# MOD derived pyrochlore films as buffer layer for all-chemical YBCO coated conductors

Andrea Augieri, Angelo Vannozzi, Rita Mancini, Achille Armenio Angrisani, Fabio Fabbri, Valentina Galluzzi, Alessandro Rufoloni, Francesco Rizzo, Antonella Mancini, Giuseppe Celentano, Ivan Colantoni, Ivan Davoli, Nicola Pompeo, Giovanni Sotgiu, Enrico Silva

*Abstract*— We report a detailed study performed on La$_2$Zr$_2$O$_7$ (LZO) pyrochlore material grown by Metal-Organic Decomposition (MOD) method as buffer layers for YBa$_2$Cu$_3$O$_{7-x}$ (YBCO) coated conductors. High quality epitaxial LZO thin films have been obtained on single crystal (SC) and Ni-5%at.W substrates. In order to evaluate structural and morphological properties, films have been characterized by means of X-ray diffraction analyses (XRD), atomic force microscope (AFM) and scanning electron microscope (SEM). Precursors solutions and heat treatments have been studied by thermogravimetric analyses (TG-DTA-DTG) and infrared spectra (FT-IR) with the aim of optimizing the annealing process. Thin films of YBCO have been deposited by pulsed laser ablation (PLD) on this buffer layers. The best results obtained on SC showed YBCO films with critical temperature values above 90 K, high self field critical current density values ($J_c > 1$ MA/cm$^2$) and high irreversibility field values (8.3 T) at 77 K together with a rather high depinning frequency $\nu_p$ (0.5 T, 77 K)>44 GHz as determined at microwaves. The best results on Ni-5%at.W has been obtained introducing in the heat treatment a pyrolysis process at low temperature in air in order to remove the residual organic part of the precursor solution.

## I. INTRODUCTION

YBCO COATED CONDUCTORS with transport performances meeting the requirements of many applications are nowadays commercially available [1]. However, in order to reach a widespread use of coated conductors in applications there are still many scientific and technological issues to be solved [2,3]. In particular, one of the most important features for the market is the cost per Ampere meter parameter (€/kAm), which sets the ratio between costs and performances

Manuscript received October 9, 2012. This work was supported by MIUR in the framework of the FIRB-Futuro in Ricerca project "SURE:ARTYST".

Andrea Augieri. Achille Armenio Angrisani, Valentina Galluzzi, Fabio Fabbri, Angelo Vannozzi, Alessanfro Rufoloni, Francesco Rizzo, Antonella Mancini, Giuseppe Celentano are with ENEA, Via Enrico Fermi 45, 00044 Frascati, Rome (Italy) (phone: +390694005994; fax: +390694006119; e-mail: andrea.augieri@enea.it.

Rita Mancini is with ENEA, Via Anguillarese 301, 00123 S.Maria di Galeria, Rome, (Italy)

Ivan Colantoni and Ivan Davoli are with University of Rome Tor Vergata, Physics Department, Via della Ricerca Scientifica 1, 00133 Rome (Italy).

Nicola Pompeo and Enrico Silva are with University of Roma Tre, Physics Department, Via della Vasca Navale 84, 00146 Rome (Italy).

Giovanni Sotgiu is with University of Roma Tre, Engineering Department, Via della Vasca Navale 79, 00146 Rome (Italy).

of the tape. Chemical Solution Deposition (CSD) techniques for RABiTS (Rolling Assisted Bi-axially Textured Substrates) based technology represent a good chance to reduce this parameter for long length coated conductor production. These techniques, in fact, don't rely with expensive deposition systems like standard vapor physical deposition methods such as Pulsed Laser Deposition (PLD) or hybrid methods such as Metal-Organic Chemical Vapor Deposition (MOCVD) [4-6]. Moreover, chemical methods are characterized by a high degree of scalability and a precise control of precursors stoichiometry. Among the materials proposed as buffer layers, La$_2$Zr$_2$O$_7$ (LZO) represents one of the best candidates because of its structural compatibility and low lattice mismatch with YBCO (≤1.8%) and Ni-5at.%W tapes (7.6%). Many groups [7-12] have recently demonstrated the possibility to grow high quality LZO films through CSD methods. LZO films have also been validated as buffer layer for MOCVD or PLD YBCO [13-15]. However, one of the most critical issues concerning CSD methods is the film surface quality. CSD-LZO films show sponge-like surfaces [16] influencing the ability of the layer to act as oxygen and metallic ions diffusion barrier. Moreover, surface quality of buffer layers is a critical parameter for the deposition of MOD YBCO in all-chemical coated-conductor structures.

In this work, we report on a detailed study performed on LZO films deposited by spin coating on both single crystals and metallic tapes starting from pentanedionate precursors. Our results show that the introduction in the LZO film growth process of a pyrolysis step performed at low temperature in air is effective in improving LZO surface quality.

## II. EXPERIMENTAL

LZO precursor solutions have been prepared by dispersing a stoichiometric mixture of La and Zr Pentanedionate in an excess of propionic acid. The as obtained solution has been treated in an ultrasonic bath at 40 °C for 20 minutes and then rotoevaporated under severe conditions (75 °C - 35 mbar) until the achievement of the desired concentration. Part of the solution has been powdered in air at 120 °C for several hours. The as obtained powder has been characterized by Thermogravimetric (TG-DTA-DTG) analyses and Fourier Transform Infrared Spectroscopy (FT-IR).

Spin coated LZO thin films have been grown on both SrTiO$_3$ single crystal and Ni-5at.%W metallic tapes prepared at ENEA or provided by EVICO Gmbh. The rotation speed has been set to 3000 RPM for 60 s. After the spinning, films have been dried for 20 minutes at 120 °C in air prior to the heat treatment. The annealing has been performed with a 10°C/min ramp. Background atmosphere (air or Ar/H₂) and annealing temperatures varied for each sample.

LZO films have been characterized by X-ray diffraction, Scanning Electron Microscope (SEM) and Atomic Force Microscope (AFM) in order to optimize their structural, morphological and texture properties. On top of LZO layer a thin film of YBCO has been deposited by PLD with deposition parameters optimized for YBCO on single crystal and reported elsewhere [17].

D.C. superconductive properties have been measured in the four-probe configuration. Patterned samples (50 μm wide 1 mm long stripes) have been mounted in a He flow cryostat provided with a 12 T superconducting magnet in helium bath. A criterion of 1μVcm⁻¹ has been used to extract critical current value from the current-voltage characteristics recorded as a function of temperature and applied magnetic field intensity and direction (by keeping the maximum Lorentz force configuration).

### III. RESULTS AND DISCUSSION

In order to reproduce the typical heat treatment used for LZO films on metallic tapes, DTA-TG analyses have been performed on powdered LZO solution in flowing (1.5 s.c.f.h.) mixture Ar + 5%H₂ with a ramp rate of 10 °C/min (figure 1).

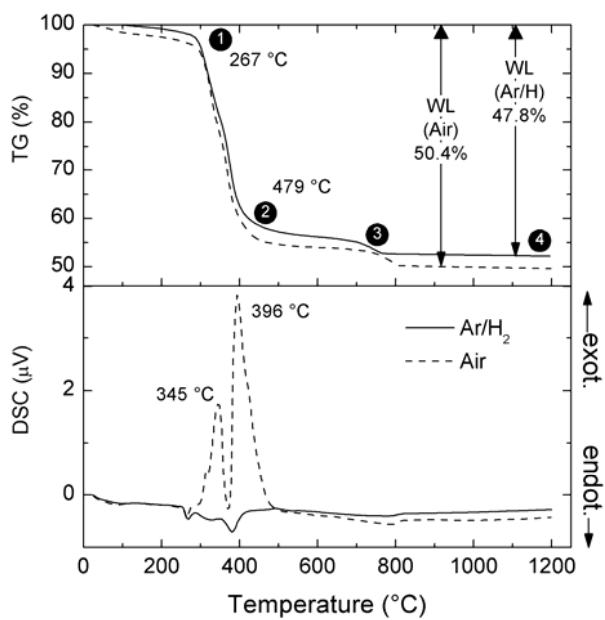

Fig. 1. DSC and TG curves of powdered LZO soultion acquired in flowing Ar/H₂ (continous lines) and Air (dashed lines) atmospheres with a ramp rate of 10 °C/min. Numbers in the upper panel identify the LZO decomposition steps.

As previously reported by other groups [8], the decomposition process of LZO can be divided into 4 steps. The first step is characterized by a weight loss of 2.4%, which can be ascribed to the evaporation of some residual water in powders. Almost the whole precursor solution decomposition occurs during the second step, which is characterized by a huge weight loss. According to previously reported studies [18, 19], in this range of temperatures almost all the Zr should be decomposed into ZrO₂ while La into La₂O(CO₃)₂. The decomposition of La-oxy-carbonate and the formation of the LZO phase are completed in the last two steps. The total weight loss measured in the process is 47.8%. In order to establish whether the reducing atmosphere influences the decomposition process of LZO powders, the same heat treatment has been repeated in flowing air. The results of DTA-TG analyses are shown in figure 1 and reveal a completely different thermal response. Two exothermic peaks at 345 °C and 396 °C are clearly noticeable, which can be ascribed to carbon combustion. As a result, the total weight loss measured is increased to 50.4%. It is worth noting that the weight loss value measured on the powders annealed in Ar/H, increases to 50.1 % if the treated powders are re-annealed in Air, thus confirming the role of oxygen in promoting carbon removal.

However, the measured values are all lower than 61.4 %, which is the expected value for the LZO decomposition process starting from La and Zr propionates. The difference between the theoretical value and the measured values can be ascribed both to a different composition of the starting powders and to the presence of a residual organic component into the annealed powders. The presence of unreacted La and Zr acetates into the starting powder can be ruled out by FTIR spectra performed on the powdered solution (not shown here) showing the absence of the characteristic vibrations of acetates and the presence of other bands which are usually ascribed to propionates. For the same reason, the absence of the characteristic band correlated to the O-H group vibrations suggests that the presence of propionate hydroxide can be considered as negligible. On the other hand, X-ray spectra and FTIR spectra performed on powdered solution rule out the presence of Zr as acetato-propionate complex ([Zr12]), as previously proposed by other groups [20]. Though the exact composition of the starting powders is unknown, these results suggest that the measured weight loss values can be mainly ascribed to the presence of a residual organic component into the treated powders. Assuming that the difference between the theoretical and the measured weight loss values is due to unreacted carbon, the amount of carbon can be estimated in 6.2 moles each La mole. This result, though unexpected, is in agreement with ICP mass measurement results obtained by other groups [9].

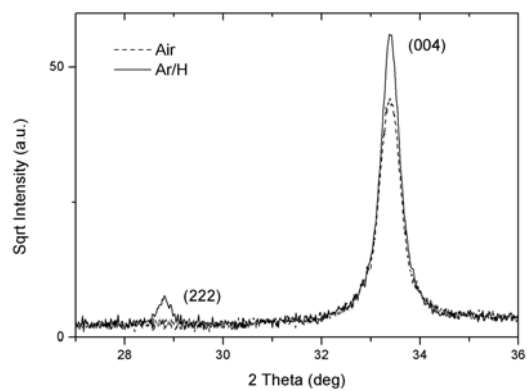

Fig. 2. X-ray diffraction spectra in $\theta$-$2\theta$ configuration showing the (004) component and (222) component of LZO films.

With 0.16 M concentrated (with respect to La ion) coating solution, 40 nm thick LZO films have been deposited by spin coating method on SrTiO₃ single crystal substrates. The annealing temperature has been fixed at 950 °C for 30 minutes. Both flowing Ar/H₂ and air have been used as treatment atmospheres. X-ray spectra performed on as grown films, revealed that in both cases only the LZO phase can be detected. As shown in figure 2, sample annealed in air exhibit only (004) peak corresponding to the right $c$-axis orientation while for the sample produced in Ar/H₂ flow also the (222) component is clearly visible. This feature can be ascribed to residual carbon which promotes undesired orientations, although the influence of residual organic component on the film growth has not been clearly understood. Taking the peak intensity ratio between $c$-axis and (222) components, $\beta$, as a measure of the film orientation quality, we evaluated $\beta$=0.99 and $\beta$=0.95 respectively for sample annealed in air and Ar/H₂. These values indicate that, in both cases, films are highly $c$-axis oriented and suitable as YBCO deposition template.

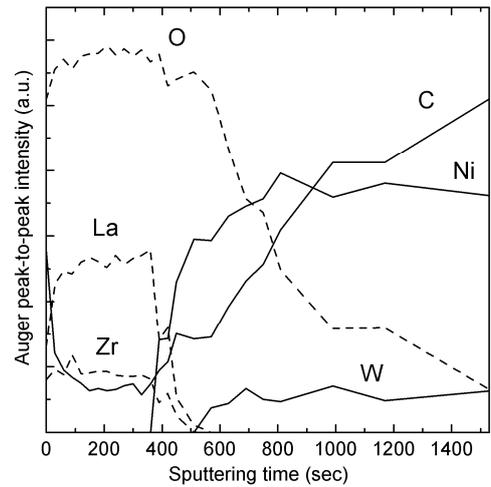

Fig. 4. Auger depth profile for a LZO film grown on Ni-5at.%W tape subjected to thermal treatment simulating YBCO PLD deposition. Signals intensity ratio does not represent elements relative concentration.

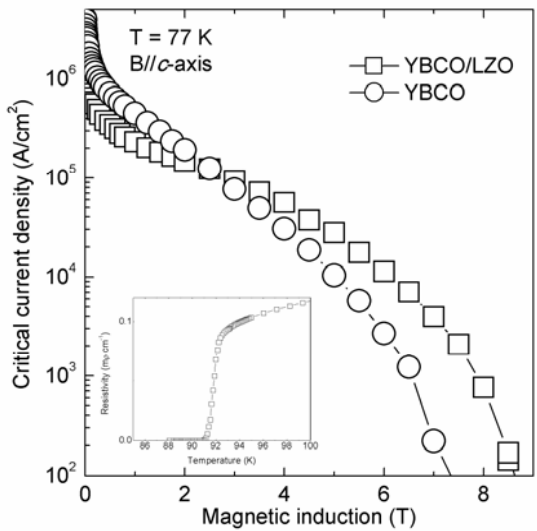

Fig. 3. Critical current density as a function of applied magnetic field ($B$ parallel to film $c$-axis) recorded at T = 77 K in PLD YBCO grown on MOD LZO buffered STO (squares) and PLD YBCO grown directly on STO (circles). In the inset the superconductive transition of YBCO/LZO film is showed.

A 200 nm thick YBCO film has been deposited by PLD on LZO film annealed in air, showing good critical temperature values as high as 91 K and good D.C. transport properties. As shown in figure 3, a better in field retention of the critical current density ($J_c$) and a higher irreversibility field value ($H_{irr}$ = 8.3 T) if compared with standard YBCO/STO have been measured at $T$ = 77 K with magnetic field oriented along the film $c$-axis. Preliminary results of $J_c$ measurements as a function of applied field intensity and direction, suggest the presence of correlated defects along the YBCO/LZO $c$-axis direction, which can be responsible for the observed transport properties. This result has been also confirmed by microwave measurements [21] performed on the same samples, revealing a pinning as strong as in YBCO with BaZrO₃ nano-inclusions.

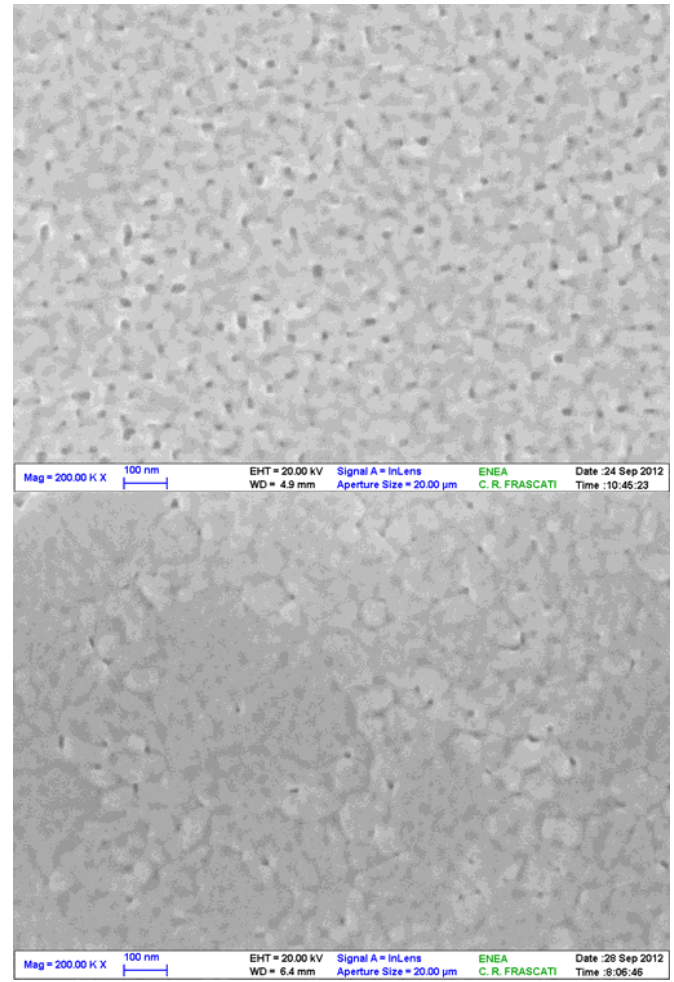

Fig. 5. Scanning Electron Microscope (SEM) images of LZO films surface grown on Ni-5at.%W metallic tapes. Upper panel: film grown with standard heat treatment; lower panel: film grown with two step heat treatment (see text for details).

A more concentrated LZO coating solution (0.35 M) has been used to deposit 120 nm thick films on Ni-5at.%W tapes with a single spin coating process. In this case the annealing temperature has been set to 1050 °C for 40 minutes and the heat treatment has been performed in flowing (1.5 s.c.f.h.) Ar/H₂. Films obtained with this process showed a high degree of $c$-axis orientation ($\beta$=0.99) and crack-free surfaces. In order to evaluate the effectiveness in blocking metal diffusion from the tape, samples have been analyzed by means of Auger spectroscopy depth profiling before and after an annealing simulating PLD-YBCO deposition (850 °C in 300 mTorr oxygen atmosphere). The results obtained on as grown films showed the presence of a small inter-diffusion region (about 20 nm) where Ni diffuses inside LZO film. After the annealing (see figure 4), the inter-diffusion region remains unaltered, while deep oxygen diffusion into the tape can be clearly seen. The presence of Ni at tape surface cannot be detected, thus confirming that 120 nm thick LZO film acts effectively as Ni diffusion barrier, in agreement with the results obtained with similar techniques by other groups [22]. It is worth noting that Auger spectroscopy revealed the presence of residual carbon into the LZO film even after the annealing simulating YBCO deposition.

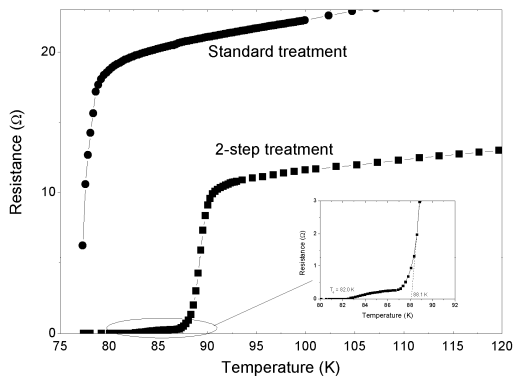

Fig. 6. Superconductive resistive transition measured on PLD YBCO films grown on LZO buffered Ni-5at.%W tapes treated with the standard annealing (circles) and the modified two steps annealing (squares). In the inset a magnification of the transition region for the sample obtained with the two steps annealing is shown.

SEM investigations shown in figure 5 (upper panel), reveal that films exhibit a porous surface, characterized by a presence of 10 to 30 nm sized voids with a density $n_{n-v} \sim 200$ μm⁻². The presence of such nano-voids on the LZO film surface (and also inside the film) have been reported by several groups [16] and is usually ascribed to the evaporation of the residual organic component inside the unconverted film. A 200 nm thick YBCO film has been deposited by PLD on the LZO film, but YBCO showed poor superconductive properties. As shown in figure 6, the resistive transition was characterized by a low onset temperature (about 80 K) and incomplete transition to the superconductive phase down to 77 K. The study of YBCO film surface morphology performed by

SEM analyses reveals a sponge-like structure composed by 100 to 200 nm holes which can be correlated to the high density on nano-voids observed on LZO surface. This is a direct evidence that the presence of residual carbon into LZO films affects the film texture and morphology with severe drawbacks on YBCO properties. For this reason a modified heat treatment was developed for LZO film growth with a low temperature pyrolysis step performed in air. The temperature of the pyrolysis step was set at 350 °C, high enough to promote carbon combustion without compromising the metallic tape through oxidation. Pyrolysis is followed by the standard annealing at 1050 °C in flowing Ar/H₂. As shown in the lower panel of figure 5, the surface morphology of LZO film obtained with the modified heat treatment is different. The surface is characterized by a very low density of nano-voids ($n_{n-v} \sim 22$ μm⁻²) and by the presence of big islands (300 to 400 nm diameter) due to the coalescence of smaller grains. The surface is very compact and smooth as revealed by AFM analyses which evaluated the root mean square value for the roughness in a 2x2 μm area in $R_{rms} = 3.5$ nm ($R_{rms} = 6.5$ nm for LZO films obtained with standard heat treatment). YBCO film deposited by PLD method on this LZO film showed improved superconductive properties, with high transition onset temperature ($T_{c,onset} > 90$ K). The measured value for the critical temperature $T_c = 82$K, does not completely reflect the improvement obtained with the two step heat treatment. As shown in the inset of figure 6, at least 90% of the transition was completed at 88 K. These promising results indicate that LZO surface, and consequently, YBCO properties can be improved by optimizing the low temperature pyrolysis process.

## IV. Conclusions

We reported on spin coated LZO films grown on STO single crystals and Ni-5at.%W metallic tapes starting from coating solution with La and Zr pentanedionate. DTA-TG characterizations performed on powdered solution revealed a large amount of residual carbon in treated solution. Residual carbon strongly influences the microstructural and texture quality of buffer layers both on single crystal and metallic tape. For LZO film on STO, the heat treatment performed in oxidizing atmosphere promotes carbon removal thus allowing high quality buffer layers. PLD YBCO grown on these films exhibits very good superconductive properties. On the contrary, LZO films grown on tape show poor surface morphology due to the reducing atmosphere used during the heat treatment. These features directly affect superconductive properties of PLD YBCO final layer. The use of a pyrolysis treatment performed in air followed by standard annealing in Ar/H₂ improves buffer quality with direct positive consequences on YBCO properties.

## Acknowledgment

This work has been founded by MIUR in the framework of the FIRB-Futuro in Ricerca project SURE:ARTYST.